\documentclass[twocolumn,pra,article,showpacs]{revtex4-1}
\usepackage{bm}
\usepackage{epsfig}
\usepackage{graphicx}
\usepackage{amssymb}
\usepackage{amsthm}
\usepackage{amsmath}
\newcommand{\F}{\hat{\mathbf{F}}}
\newcommand{\Fx}{\hat{F}_x}
\newcommand{\Fy}{\hat{F}_y}
\newcommand{\Fz}{\hat{F}_z}

\newcommand{\Hthree}{\hat{B}_{2;-1}^3}
\newcommand{\Hthreep}{\hat{B}_{2;-1}^{3'}}
\newcommand{\Hfour}{\hat{B}_{2;-1}^{4}}

\newcommand{\f}{\mathbf{f}}
\newcommand{\U}{\hat{U}}
\newcommand{\X}{\hat{X}}
\newcommand{\Y}{\hat{Y}}
\newcommand{\beq}{\begin{equation}}
\newcommand{\enq}{\end{equation}}

\begin{document}

\title{Stability of nonstationary states of spin-$2$ Bose-Einstein condensates}

\author{H. M\"akel\"a and E. Lundh}
\affiliation{Department of Physics, Ume\aa \,\,University, SE-901 87 Ume\aa, Sweden}
%\date{September 8, 2011}

\begin{abstract}
The dynamical 
stability of nonstationary states of homogeneous spin-2 rubidium Bose-Einstein 
condensates is studied.  
The states considered are such that the spin vector remains parallel to the magnetic field 
throughout the time evolution, making it possible to study the stability analytically. These  states are shown to be 
stable in the absence of an external magnetic field, but they become unstable when a finite magnetic field is introduced. It is found that the growth rate and wavelength of the instabilities can be controlled by tuning the strength of the magnetic field and the size of the condensate.
\end{abstract}


\maketitle

\section{Introduction}
The physics of $F=2$ spinor Bose-Einstein condensates (BECs) started to gain the attention of both theorists and experimentalists during the last decade. The interest was motivated by the structure of $F=2$ condensates: being more complex than that of $F=1$ condensates, it made  possible properties and phenomena which are not present in an $F=1$ system.
One example of this can been seen in the structure of the ground states.  
The energy functional of an $F=2$ condensate is characterized by one additional degree of freedom compared to the $F=1$ case.  
This leads to a rich ground state manifold   
as now there are two free parameters parametrizing the ground states \cite{Ciobanu00,Zheng10}. 
This should be contrasted with an $F=1$ condensate, where the ground state is determined by the sign 
of the spin-dependent interaction term \cite{Ho98,Ohmi98}.  
Another difference can be seen in the structure of topological defects. It has been shown that non-commuting vortices can exist in an $F=2$ condensate \cite{Makela03}, while these are not possible in an $F=1$ BEC \cite{Ho98,Makela03}. The topological defects of $F=2$ condensates have been studied further by the authors of Refs. \cite{Makela06,Huhtamaki09,Kobayashi09}.
Experimental studies of $F=2$ BECs have been advancing in the past ten years.
Experiments on $F=2$ ${}^{87}$Rb atoms 
cover topics such as spin dynamics \cite{Schmaljohann04,Chang04,Kuwamoto04,Kronjager06,Kronjager10}, creation of skyrmions \cite{Leslie09a}, spin-dependent inelastic collisions \cite{Tojo09}, amplification of fluctuations \cite{Klempt09,Klempt10}, spontaneous breaking of spatial and spin symmetry \cite{Scherer10}, and atomic homodyne detection \cite{Gross11}. An $F=2$ spinor condensate of ${}^{23}$Na atoms has been obtained experimentally \cite{Gorlitz03}, but it has a much shorter lifetime than $F=2$ rubidium condensates.

In this work, we study the dynamical stability of nonstationary states of homogeneous 
$F=2$ spinor condensates. The stability of stationary states has been examined 
both experimentally \cite{Klempt09,Klempt10,Scherer10} and theoretically \cite{Martikainen01,Ueda02}. 
Interestingly, the experimental studies show that the observed instability of the $|m_F=0\rangle$ state can be used to amplify vacuum fluctuations \cite{Klempt10} and to analyze symmetry breaking \cite{Scherer10} 
(see Refs. \cite{Lamacraft07,Leslie09b} for related studies in an $F=1$ system).
The stability of nonstationary states of spinor condensates, on the other hand, has received only little attention. Previous studies on the topic  concentrate on $F=1$ condensates  \cite{Matuszewski08,Matuszewski09,Matuszewski10,Zhang05,Makela11}. 
Here we extend the analysis of the authors of Ref. \cite{Makela11} to an $F=2$ rubidium condensate and present results 
concerning the magnetic field dependence of the excitation spectrum and stability. 
Although we concentrate on the stability of ${}^{87}$Rb condensates, many of the excitation spectra and stability conditions given in this article are not specific to rubidium condensates but have a wider applicability. 
We show that, in comparison with an $F=1$ system, the stability analysis of an $F=2$ condensate is considerably 
more complicated. This is partly due to the presence of a spin-singlet term in the energy functional of the latter system, but the main reason for the increased complexity is seen to be the much larger number of states available in an $F=2$ condensate.

This article is organized as follows. Section \ref{sec:overview} introduces the system and presents the Hamiltonian and equations of motion. In Sec. \ref{sec:stability} the Bogoliubov analysis of nonstationary states is introduced. This method is applied to study the stability both in the presence and absence of a magnetic field. 
In this section it is also described how Floquet theory can be used in the stability analysis. 
In Sec. \ref{sec:g2not0} the stability is studied under the (physically motivated) assumption that one of the interaction coefficients vanishes. Finally, Sec. \ref{sec:conclusions} contains the concluding remarks.

\section{Theory of a spin-2 condensate}\label{sec:overview}

The order parameter of a spin-$2$ Bose-Einstein condensate can be written as $\psi=(\psi_2,\psi_{1},\psi_{0},\psi_{-1},\psi_{-2})^T$, where $T$ denotes the transpose. The normalization is $\sum_{m=-2}^2|\psi_{m}|^2=n$, where $n$ is the total particle density. We assume that the trap confining the condensate is such that all the
components of the hyperfine spin can be trapped simultaneously and are degenerate 
in the absence of magnetic field. This can be readily achieved in experiments \cite{Stamper-Kurn98}.   
If the system is exposed to an external magnetic field which 
is parallel to the $z$ axis, the energy functional reads 
\begin{align}
\label{energy}
&E[\psi] =\!\! \int d{\bm r} 
\left[ \langle \hat{h}\rangle  +\frac{1}{2}\left(g_0 n^2  + g_1 \langle\F\rangle^2 
 + g_2 |\Theta|^2\right)\right],
\end{align}
where $\F=(\Fx,\Fy,\Fz)$ is the (dimensionless) spin operator of a spin-2 particle. 
 $\Theta$ describes singlet pairs and is given by $\Theta=2\psi_2\psi_{-2}-2\psi_1\psi_{-1}+\psi_0^2$. It can also be written as $\Theta=\psi^T e^{-i\pi \hat{F}_y}\psi$. The single-particle Hamiltonian $\hat{h}$ reads  
\begin{align}
\label{h}
\hat{h}= -\frac{\hbar^2 \nabla^2}{2m} + U(\mathbf{r})  -\mu-p\Fz+q\Fz^2. 
\end{align} 
Here $U$ is the external trapping potential, $\mu$ is the chemical potential, and  
$p=-g\mu_{\rm B}B$ is the linear Zeeman term. In the last of these  $g$ is the Land\'e hyperfine $g$-factor, $\mu_{\rm B}$ is the Bohr magneton, and $B$ is the external magnetic field. 
The last term in Eq. (\ref{h}) is  the quadratic Zeeman term, $q=-(g\mu_{\rm B}B)^2/E_{\rm hf}$,   where $E_{\rm hf}$ is the hyperfine splitting. The sign of $q$ can be controlled experimentally   by using a linearly polarized microwave field \cite{Gerbier06}. In this article we consider both 
 positive and negative values of $q$.

The strength of the spin-independent interaction is characterized by $g_0=4\pi \hbar^2(4a_2+3a_4)/7m$, whereas $g_1=4\pi \hbar^2(a_4-a_2)/7m$ and $g_2=4\pi\hbar^2[(a_0-a_4)/5-2(a_2-a_4)/7]$ describe spin-dependent scattering.  Here $a_F$ is the $s$-wave scattering length for two atoms colliding with total angular momentum $F$. 
In the case of $^{87}$Rb, we calculate $g_0$ using the scattering lengths given in Ref.\ \cite{Ciobanu00}, and $g_2$ and $g_4$ are calculated using the experimentally measured scattering length differences from Ref.\ \cite{Widera06}.

Two important quantities characterizing the state $\psi$ are the spin vector 
\begin{align}
%\label{Mz}
\f(\mathbf{r})=  \frac{\psi^\dag(\mathbf{r}) \F \psi(\mathbf{r})}{n(\mathbf{r})},
\end{align}
and the magnetization in the direction of the magnetic field
\begin{align}
\label{Mz}
M_z= \frac{\int d\mathbf{r}\,n(\mathbf{r}) f_z (\mathbf{r})}{\int d\mathbf{r}\,n(\mathbf{r})}.
\end{align}
The length of $\f$ is denoted by $f$.
For rubidium the magnetic dipole-dipole interaction is weak and consequently the magnetization 
is a conserved quantity. The Lagrange multiplier related to the conservation of magnetization  
can be included into $p$.
The time evolution equation obtained from Eq. (\ref{energy}) is   
\begin{align}
i\hbar \frac{\partial }{\partial t}\psi =\hat{H}[\psi] \psi,
\end{align}
where
\begin{align}
\label{H}
\hat{H}[\psi]= \hat{h}+ g_0 \psi^\dag\psi +g_1 \langle\F\rangle\cdot\F + g_2 \Theta \hat{{\mathcal{T}}}. 
\end{align}
Here $\hat{{\mathcal{T}}}=e^{-i\pi \hat{F}_y}\hat{C}$ is the time-reversal operator, where $\hat{C}$ is the
complex conjugation operator.

\section{Stability of nonstationary states when $g_2\not=0$}\label{sec:stability}
The stability analysis is performed in a basis where the state in question 
is time independent. This requires that the time evolution operator of the state 
is known. As we are interested in analytical calculations, an analytical expression for  this operator has to be known. To calculate the time evolution operator analytically, the Hamiltonian has to be time independent. In particular, the singlet term $\Theta$ should not depend on time. This is clearly the case if the time evolution of the state is such that $\Theta$ vanishes at all times, and we now study this case. 
We define a state  
\begin{align}
\label{eq:psiparallel}
\psi_{2;-1}=
\sqrt{\frac{n}{3}}
\begin{pmatrix}
\sqrt{1+f_z}\\
0\\
0\\
\sqrt{2- f_z}\\
0
\end{pmatrix},\quad -1\leq f_z\leq 2.
\end{align}
For this state $\Theta=0$, $\langle \Fx\rangle=\langle \Fy\rangle=0$, and $\langle\Fz\rangle =f_z$. Furthermore, the populations of the state $\psi_{2;-1}$ remain unchanged during the time evolution 
determined by the Hamiltonian (\ref{H}). Consequently, $\Theta=0$ throughout the time evolution. 
The state $\psi_{2;-1}$ with $f_z=0$, called the cyclic state, is a ground state at zero magnetic field \cite{Ciobanu00}. The creation of vortices with fractional winding number in states of the form $\psi_{2;-1}$ has been discussed by the authors of Ref. \cite{Huhtamaki09}. The stability properties of the state 
$\psi_{1;-2}=\sqrt{n}(0,\sqrt{2+f_z},0,0,\sqrt{1-f_z})^T/\sqrt{3}$ are similar to those of $\psi_{2;-1}$ and will therefore  not be studied separately. 

The Hamiltonian giving the time evolution of $\psi_{2;-1}$ is
\begin{align}
\label{eq:Hparallel}
\hat{H}[\psi_{2;-1}]=g_0 n-\mu +(g_1 n f_z - p)\Fz +q\Fz^2,
\end{align}
where we have set $U=0$ as the system is assumed to be homogeneous. 
This is of the same form as the Hamiltonian of an $F=1$ system discussed  by the authors of Ref. \cite{Makela11}. 
The time evolution operator of $\psi_{2;-1}$ is given by 
\begin{align}
\label{eq:Uparallel}
\hat{U}_{2;-1}(t)= e^{-i t \hat{H}[\psi_{2;-1}]/\hbar}.
\end{align}

We analyze the stability in a basis where the state $\psi_{2;-1}$ is time independent. 
In the new basis, the energy of an arbitrary state $\phi$ is given by \cite{Makela11} 
\begin{align}
\label{Erot}
E^{\textrm{new}}[\phi]&=E[\U_{2;-1}\phi]+i\hbar\langle\phi|\left(\frac{\partial}{\partial t}\U^{-1}_{2;-1}\right)\U_{2;-1}\phi\rangle, 
\end{align}
and the time evolution of the components of $\phi$ can be obtained from the equation 
\begin{align}\label{variE}
i\hbar\frac{\partial\phi_\nu}{\partial t}=\frac{\delta E^{\textrm{new}}[\phi]}{\delta\phi_\nu^*},\quad \nu =-2,-1,0,1,2. 
\end{align}
We replace $\phi\rightarrow \psi_{2;-1} +\delta\psi$ in the time evolution equation (\ref{variE}) and expand the resulting equations to first order in $\delta\psi$. The perturbation $\delta\psi=(\delta\psi_2,\delta\psi_1,\delta\psi_0,\delta\psi_{-1},\delta\psi_{-2})^T$ is written as 
\begin{align*}
\delta\psi_j=\sum_{\mathbf{k}} \left[ u_{j;\mathbf{k}}(t)\,e^{i\mathbf{k}\cdot\mathbf{r}}
-v_{j;\mathbf{k}}^{*}(t)\, e^{-i\mathbf{k}\cdot\mathbf{r}}\right],
\end{align*}
where $j=-2,-1,0,1,2$. 
Straightforward calculation gives the differential equation for the time evolution of the  perturbations as
\begin{align}
i\hbar\frac{\partial}{\partial t}
\begin{pmatrix}
\mathbf{u}_\mathbf{k}\\
\mathbf{v}_\mathbf{k} 
\end{pmatrix}
&=\hat{B}_{2;-1}
\begin{pmatrix}
\mathbf{u}_\mathbf{k}\\
\mathbf{v}_\mathbf{k}
\end{pmatrix},\\ 
\label{HBG}
\hat{B}_{2;-1} &=\begin{pmatrix}
\X &- \Y\\
\Y^* & -\X^*
\end{pmatrix},
\end{align}
where 
$\mathbf{u}_\mathbf{k}=(u_{2;\mathbf{k}},u_{1;\mathbf{k}},u_{0;\mathbf{k}},u_{-1;\mathbf{k}},
u_{-2;\mathbf{k}})^T$, $\mathbf{v}_\mathbf{k}$ is defined similarly, and the $5\times 5$ matrices $\X$ and $\Y$ are
\begin{align}
\nonumber
\X =&\,\, \epsilon_k + g_0 |\psi_{2;-1}\rangle\langle\psi_{2;-1}|
+g_1 \!\!\!\!\sum_{j=x,y,z} |\psi_{2;-1}^j (t)\rangle\langle\psi_{2;-1}^j (t)| \\
\label{X}
&+2g_2  |\psi_{2;-1}^\textrm{s} (t)\rangle\langle\psi_{2;-1}^\textrm{s} (t)|, \\
\label{Y}
\Y =&\,\, g_0 |\psi_{2;-1}\rangle\langle\psi^*_{2;-1}| 
+g_1 \!\!\!\!\sum_{j=x,y,z} |\psi_{2;-1}^j (t)\rangle\langle (\psi_{2;-1}^{j})^*(t)|.
\end{align}
Here we have defined 
\begin{align}
\label{epsilonk}
\epsilon_k &= \, \frac{\hbar^2 k^2}{2m},\\
\psi_{2;-1}^j (t) &= \, \U_{2;-1}^\dag(t)\hat{F}_j U_{2;-1} (t)\psi_{2;-1},\quad j=x,y,z,\\
\psi_{2;-1}^\textrm{s} (t) &= \, \U_{2;-1}^T(t)e^{-i\pi \hat{F}_y} U_{2;-1} (t)\psi_{2;-1}. 
\end{align}
In the rest of the article we call the operator determining the time evolution of the perturbations the Bogoliubov matrix. In the present case, $\hat{B}_{2;-1}$ is the Bogolibov matrix of $\psi_{2;-1}$.  
It is possible to write $\hat{B}_{2;-1}$ as a direct sum of three operators 
\begin{align}
\hat{B}_{2;-1} (t) &=\Hfour\oplus \Hthree(t) \oplus \Hthreep(t),\\
\Hthreep &=-(\Hthree)^*. 
\end{align} 
Here $\Hfour$ is a time independent $4\times 4$ matrix and $\Hthree$ 
is a time-dependent $3\times 3$ matrix. The bases in which these operators are defined are given in Appendix \ref{sec:appendixa}. 
The time-dependent terms of $\Hthree$ are proportional to  
$e^{\pm i k q t/\hbar}$, where $k=2,4$, or $6$, and consequently the system is periodic with minimum period $T=\pi\hbar/q$. Hence it is possible to use Floquet theory to analyze the stability of the system \cite{Makela11}. In the following we first calculate the eigenvalues of $\Hfour$, then those of $\Hthree$ and $ \Hthreep$ in the case $q=0$, and finally we discuss the general case $q\not =0$ using  Floquet theory. 

\subsection{Eigenvalues of $\Hfour$}
First we calculate the eigenvalues and eigenvectors of $\Hfour$. This 
operator is independent of $q$. The eigenvalues are 
\begin{align}
\nonumber
\label{psi2m1omega1234}
\hbar\omega_{1,2,3,4} &=\pm\Big[\epsilon_k\Big(\epsilon_k +g_0 n+g_1 n(2+f_z)\\
&\pm n\sqrt{[g_0-g_1 (2+f_z)]^2+4 g_0 g_1 f_z^2}\Big)\Big]^{1/2}.
\end{align}
Here we use a labeling such that $++,-+,+-$, and $--$ correspond to $\omega_1,\omega_2,\omega_3$, and $\omega_4$, respectively. Now $\omega_{1,2}$ have a non-vanishing 
imaginary part only if $g_0$ and $g_1$ are both negative, while $\omega_{3,4}$ have an imaginary component if $g_0$ and $g_1$ are not both  positive. Consequently,  these modes are stable for rubidium for which $g_0,g_1>0$.  

The eigenvectors can be calculated straightforwardly, see Appendix \ref{sec:appendixa}. 
The eigenvectors, like the eigenvalues, are independent of $g_2$. 
The perturbations corresponding to the eigenvectors of $\Hfour$ can be written as 
\begin{align}
\delta\psi^{1,2,3,4}(\mathbf{r},\mathbf{k};t) = C_{1,2,3,4}(\mathbf{r},\mathbf{k};t)\,\psi_{2;-1}, 
\end{align}  
where the $C_j$'s include all position, momentum, and time dependence. 
These change the total density of the condensate and are therefore called  density modes.

\subsection{Eigenvalues of $\Hthree$ and $\Hthreep$ at $q=0$}
In the absence of an external magnetic field $\Hthree$ is time independent. 
The eigenvalues of $\Hthreep$ can be obtained from 
those of $\Hthree$ by complex conjugating and changing the sign. For this 
reason we give only the eigenvalues of $\Hthree$:
\begin{align}
\label{psi2m1omega5}
\hbar\omega_{5} &= \epsilon_k +2 g_2 n,\\
\label{psi2m1omega67}
\hbar\omega_{6,7} &=\frac{1}{2}\Big[g_1 n f_z
\pm \sqrt{(2\epsilon_k-g_1 n f_z)^2+ 16 g_1 n\epsilon_k}\Big].
\end{align}
The eigenvalues $\hbar\omega_6$ and $\hbar\omega_7$ have a non-vanishing complex 
part if $g_1<0$. For rubidium all eigenvalues are real. 
There are two gapped excitations: at $\epsilon_k=0$ we get $\hbar\omega_5=2g_2 n$ and $\hbar\omega_{6} 
\,(\hbar\omega_{7})=g_1 nf_z$ if $g_1 f_z>0$ $(g_1 f_z<0)$.  
The eigenvectors are given in Appendix A. 
The corresponding perturbations become
\begin{align}
\label{omega3}
\delta\psi^{5}(\mathbf{r},\mathbf{k};t) &= C_5\, 
\begin{pmatrix}
0\\
\sqrt{2-f_z}\\
0\\
0\\
-\sqrt{1+f_z}
\end{pmatrix},\\
\delta\psi^{6,7}(\mathbf{r},\mathbf{k};t) &= \sum_\mathbf{k} C_{6,7} \, 
\begin{pmatrix}
0\\
e^{i \mathbf{k}\cdot \mathbf{r}} g_1 n\sqrt{(2-f_z)(1+f_z)}\\ 
e^{-i \mathbf{k}\cdot \mathbf{r}}\sqrt{\frac{3}{2}}(\epsilon_k+2 g_1 n-\hbar\omega_{6,7})\\
0\\
e^{i \mathbf{k}\cdot \mathbf{r}} g_1 n (2-f_z),
\end{pmatrix}
\end{align}
where $C_{5,6,7}$ are functions of $\mathbf{r},\mathbf{k}$, and $t$.
These modes change both the direction of the spin and magnetization and are therefore called spin-magnetization modes.

\subsection{Non-vanishing magnetic field}
If $q\not =0$, the stability can be analyzed using Floquet theory due to the 
periodicity of  $\Hthree$ \cite{Makela11}. We denote the time evolution operator 
determined by $\Hthree$ by $\hat{U}_{2;-1}^3$.   
According to the Floquet theorem (see, e.g., Ref. \cite{Chicone}), $\hat{U}_{2;-1}^3$ can be written as 
\begin{align}
\U_{2;-1}^3(t)=\hat{M}(t) e^{-i t \hat{K}},
\end{align}
where $\hat{M}$ is a periodic matrix with minimum period $T$ and $\hat{M}(0)=\hat{\textrm{I}}$, 
and $\hat{K}$ is some time-independent matrix. 
At times $t=nT$, where $n$ is an integer, we get $\U_{2;-1}^3(nT)=e^{-i n T \hat{K}}$. The eigenvalues of $\hat{K}$ determine the stability of the system. 
We say that the system is unstable if at least one of the eigenvalues of $\hat{K}$ has a positive imaginary part. We calculate the eigenvalues $\{\hbar\omega\}$ of $\hat{K}$ from the equation
\begin{align}
\hbar\omega=\hbar\omega^\textrm{r}+i \hbar\omega^\textrm{i}=i\frac{\ln\lambda}{T}, 
\end{align}
 where $\{\lambda\}$ are the eigenvalues of $\U_{2;-1}^3(T)$.

\begin{figure}[h]
\begin{center}
\includegraphics[scale=.7]{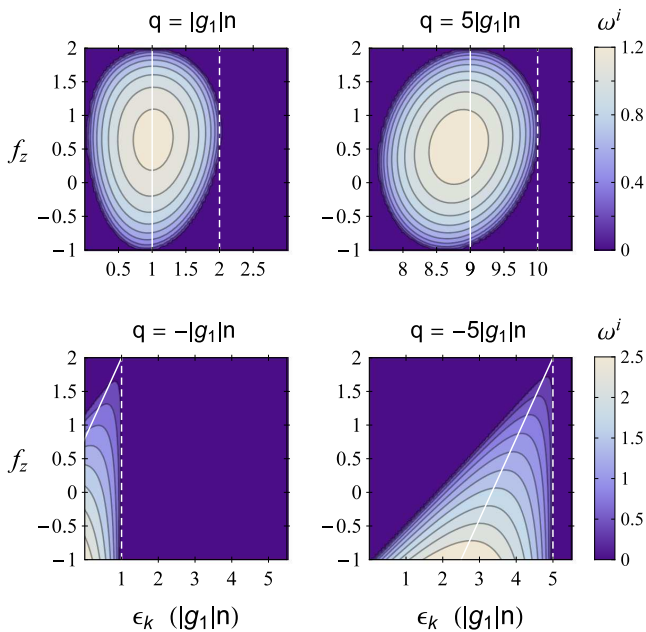} 
\end{center}
\caption{(Color online) The positive imaginary part $\omega^\textrm{i}$ related to  $\hat{U}_{2;-1}(T)$ for different values of the magnetic field parameter $q$. 
The unit of $\omega^\textrm{i}$ is $|g_1|n/\hbar$. Note that the scales of $\epsilon_k$ and $\omega^\textrm{i}$ are not equal in the top and bottom rows. Note also that the scale of the $q=5|g_1|n$ figure  
is shifted with respect to the scale of the $q=|g_1|n$ case. The solid white line gives the approximate location of the fastest-growing instability, and the dashed white line corresponds to the largest possible size of a 
stable condensate, see Eq. (\ref{lambda2m1}) and Table \ref{table}. 
\label{fig:psi2m1}}
\end{figure} 
 
We plot $\omega^\textrm{i}$ for several values of the magnetic field in Fig. \ref{fig:psi2m1}. 
By comparing this to the case of a rubidium condensate with $g_2=0$,  
we found that the instabilities are essentially determined by $g_1$, the effect of $g_2$ is negligible. 
The eigenvectors of $\hat{U}_{2;-1}^3(T)$ correspond to perturbations which affect both 
spin direction and magnetization.   
With the help of numerical results we find that a good fitting formula is given by 
\begin{align}
\hbar\omega^\textrm{i}
\approx &\textrm{Im}\Big\{
\sqrt{(\epsilon_k+q)[\epsilon_k+q+\frac{5}{3}(2-f_z) g_1 n]}\Big\}, q<0,\\
\nonumber
\approx &\textrm{Im}\Big\{\sqrt{(\epsilon_k-2q+ g_1 n)^2-\frac{4}{9}|(f_z-2)(f_z+1)g_1n|}\Big\},\\
& q>0.
\end{align}
We see that for $q>0$ the fastest-growing instability is located approximately at $\epsilon_k=\max\{0,2q-g_1 n\}$ regardless of the value of $f_z$. For $q<0$ the location 
of this instability becomes magnetization dependent and is approximately given 
by $\epsilon_k=\max \{0, |q|-5(2-f_z)|g_1|n/6\}$.  
The values of $\epsilon_k$ corresponding to unstable wavelengths are bounded above approximately  by the inequality $\epsilon_k\leq (3|q|+q)/2$. Therefore, the state $\psi_{2;-1}$ 
is stable if the condensate is smaller than the shortest unstable wavelength
\begin{align}
\lambda_{2;-1} =\frac{2\pi\hbar}{\sqrt{m(3|q|+q)}}. 
\label{lambda2m1}
\end{align}
At $q=0$ the system is stable regardless of its size.  

Figure \ref{fig:psi2m1} shows that the shape of the unstable region  
depends strongly on the sign of $q$. This can be understood qualitatively 
with the help of the energy functional of Eq. (\ref{energy}). 
We choose $\psi_{\textrm{ini}}=\sqrt{n}|m_F=-1\rangle$ to be the initial state of the system and assume that the 
final state is of the form 
\begin{align}
\psi_{\textrm{fin}}(x,y,z)=
\begin{cases}
\sqrt{n}|0\rangle, & x\, \textrm{mod}\, 2L \in [0,L),\\
\sqrt{n}|-2\rangle, & x\, \textrm{mod}\, 2L \in [L,2L).
\end{cases}
\end{align}
Then the energy of the initial state is $E_{\textrm{ini}}=g_1 n/2+q$ (dropping constant terms), 
while the energy of the final configuration   
is $E_{\textrm{fin}}=g_1 n+ 2q$. If $g_1,q>0$, $E_{\textrm{ini}}<E_{\textrm{fin}}$ and domain formation is suppressed for energetic reasons. 
If, on the other hand, $g_1>0$ and $q<0$, the energy of the final state is smaller than the energy of the initial state if $q<-g_1 n/2$ and domain formation is possible.

\section{Stability of nonstationary states when $g_2=0$}\label{sec:g2not0}
For rubidium the value of $g_2$ is small in comparison with $g_0$ and $g_1$. 
Consequently, it can be assumed that this term has only a minor effect on the stability of the system. This assumption is supported by the results of the previous section.  
In the following we will therefore study the stability in the limit $g_2=0$. 
This makes it possible to obtain an analytical expression for the time evolution 
operator also for states other than $\psi_{2;-1}$. First we discuss a state that has three nonzero components, 
and then two states that have two nonzero components.  

\subsection{Nonzero $\psi_2$, $\psi_0$, and $\psi_{-2}$} 
We consider a state of the form 
\begin{align}
\psi_{2;0;-2}=
\frac{\sqrt{n}}{2}\begin{pmatrix}
\sqrt{2-2 \rho_0+f_z}\\
0\\
2e^{i\theta} \sqrt{\rho_0}\\
0\\
\sqrt{2-2 \rho_0-f_z}
\end{pmatrix},\quad |f_z|\leq 2-2\rho_0
\end{align}
For this $\langle\hat{F}_x\rangle=\langle\hat{F}_y\rangle=0$ and 
$\langle\hat{F}_z\rangle= nf_z$. The Hamiltonian and time evolution operator of this state 
are given by Eqs. (\ref{eq:Hparallel}) and (\ref{eq:Uparallel}), respectively. 
The equations determining the time evolution of the perturbations can be obtained 
from Eqs. (\ref{X}) and (\ref{Y}) by replacing $\psi_{2;-1}$ with $\psi_{2;0;-2}$ 
and setting $g_2=0$. In this way, we obtain a time dependent Bogoliubov matrix 
$\hat{B}_{2;0;-2}$, which is a function of the population of the zero component $\rho_0$.  
The Bogoliubov matrix can now be written as
\begin{align}
\label{eq:HB20m2}
\hat{B}_{2;0;-2} (t) =\hat{B}_{2;0;-2}^6\oplus \hat{B}_{2;0;-2}^4(t), 
\end{align} 
where $\hat{B}_{2;0;-2}^6$ is time independent and 
$\hat{B}_{2;0;-2}^4$ is periodic in time with period $T=\pi/q$. 
The bases in which these operators are defined are given in Appendix B. 
The eigenvalues of $ \hat{B}_{2;0;-2}^6$ are 
\begin{align}
\hbar\omega_{1,2} =& \pm\epsilon_k,\\
\nonumber 
\hbar\omega_{3,4,5,6} =& \pm\Big[\epsilon_k^2 +\epsilon_k [g_0+4g_1(1-\rho_0)]n \\
&\pm \epsilon_k n\sqrt{[g_0-4g_1(1-\rho_0)]^2+4 g_0 g_1 f_z^2}\Big]^{1/2}.
\end{align}
Here $++,-+,+-$, and $--$ correspond to $\omega_3,\omega_4,\omega_5$, 
and $\omega_6$, respectively. 
These eigenvalues are always real if $g_0$ and $g_1$ are positive. 
From the eigenvectors given in Appendix \ref{sec:appendixb} we see that $\omega_{3,4}$ are density modes and $\omega_{1,2}$ and $\omega_{5,6}$ are magnetization modes. All these are gapless excitations. 
Note that the eigenvalues are independent of $\theta$. 

We discuss next the stability properties determined by $\hat{B}_{2;0;-2}^4$. We consider first the special case $\rho_0=0$  and proceed then to the case $\rho_0>0$. 

\subsubsection{Stability at $\rho_0=0$}
\begin{figure}[h]
\begin{center}
\includegraphics[scale=.86]{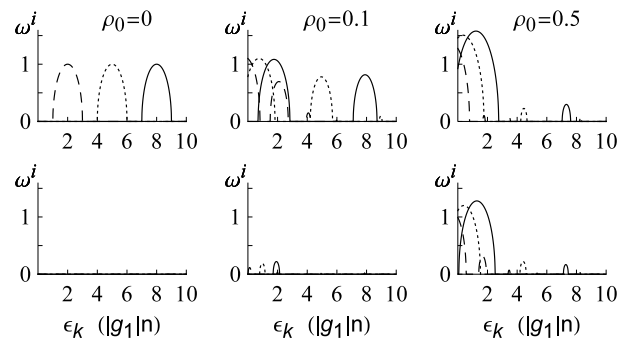} 
\end{center}
\caption{The positive imaginary part $\omega^\textrm{i}$ related to the eigenvalues (\ref{psi20m2omega78910}) and
to  $\hat{U}_{2;0;-2}(T)$  
for various values of the quadratic Zeeman term $q$ and population $\rho_0$. 
The unit of $\omega^\textrm{i}$ is $|g_1|n/\hbar$. 
We have chosen $f_z=0$ as this choice gives the fastest-growing instabilities and the smallest size of a stable condensate. In the top row the dashed, dotted, and solid lines correspond to $q=1,2,3$, respectively, while in the bottom row they correspond to $q=-1,-2,-3$, respectively. We have set $\theta=0$ in $\psi_{2;0;-2}$ as 
the stability was found to be independent of $\theta$. 
 \label{fig:psi20m2}}
\end{figure}

In the case $\rho_0=0$ a complete analytical solution of the excitation spectrum can be obtained. 
In Appendix \ref{sec:appendixb} we show that by a suitable choice of basis the time dependence of the  Bogoliubov matrix can be eliminated. The eigenvalues are 
\begin{align}
\nonumber
\label{psi20m2omega78910}
&\hbar\omega_{7,8,9,10} = \frac{1}{2}\Big[\pm g_1 n f_z + 6q \\
&\pm\sqrt{4(\epsilon_k+g_1 n-3 q)^2
-(4-f_z^2)(g_1 n)^2} \Big].  
\end{align}
These are gapped excitations and correspond to spin-magnetization modes (see Appendix \ref{sec:appendixb}).  
If $g_1>0$, these eigenvalues have a non-vanishing complex part when $3q-2g_1 n\leq \epsilon_k \leq 3q$. This is possible only if $q$ is positive. 
The location of the fastest-growing unstable mode, determined by $\epsilon_k=\max\{0,3q - g_1 n\}$, is independent of $f_z$. The maximal width of the unstable region in the $\epsilon_k$ direction, obtained at $f_z=0$,
 is $2|g_1|n$. The state is stable if the system is smaller than the size given by 
\begin{align}
\label{lambda20m2}
\lambda_{2;0;-2} =\frac{2\pi\hbar}{\sqrt{6mq}},\quad q>0. 
\end{align}
If $q<0$, the state is stable regardless of the size of the condensate.
In Fig. \ref{fig:psi20m2} we plot the positive imaginary part of the eigenvalues (\ref{psi20m2omega78910}) for various values of $q$.

\subsubsection{Stability when $\rho_0>0$}
In the case $\rho_0>0$ the stability can be studied using Floquet theory. 
The stability properties can be shown to be independent of the sign of $f_z$. 
At $q=0$ the operator $\hat{B}_{2;0;-2}^4$ is time independent. The eigenvalues 
can be obtained analytically but are not given here. 
The eigenvalues show that in the absence of magnetic field 
the state is stable in a rubidium condensate regardless of the value of $\rho_0$. 
Figure \ref{fig:psi20m2} illustrates how the stability depends on the value of $q$ and the population $\rho_0$. We plot only the case $f_z=0$ as it gives the fastest-growing 
instabilities and the smallest size of a stable condensate. 
We found numerically that the stability properties are independent of the value of $\theta$. We have set $\theta=0$ in the calculations described here.  
If $q>0$, the amplitude $\omega^\textrm{i}$ of the short-wavelength instabilities is suppressed as $\rho_0$ increases. 
This can be understood with the help of the energy functional 
\begin{align}
E_{2;0;-2} = \frac{1}{2} g_1 n f_z^2 + 8q (1-\rho_0).
\end{align}
If $q>0$, the energy decreases as $\rho_0$ increases. 
Therefore there is less energy available to be converted into the kinetic energy 
of the domain structure. From the top row of Fig. \ref{fig:psi20m2} it can be seen that Eq. (\ref{lambda20m2}) gives an upper bound for the size of a stable condensate also when the value of $\rho_0$ is larger than zero. 
If $q<0$, the state is stable at $\rho_0=0$.  The bottom row of Fig. \ref{fig:psi20m2} shows that now  $\psi_{2;0;-2}$ becomes more unstable as $\rho_0$ grows. This is natural because the energy $E_{2;0;-2}$ grows as $\rho_0$ increases, the energy surplus can be converted into kinetic energy of the domains. 
Figure \ref{fig:psi20m2} shows that Eq. (\ref{lambda20m2}) gives an upper bound for the size of a stable condensate also in the case $q<0$.  

\subsection{Nonzero $\psi_1$ and $\psi_{-1}$}
As the next example we consider a state of the form 
\begin{align}
\psi_{1;-1}=
\sqrt{\frac{n}{2}}\begin{pmatrix}
0\\
\sqrt{1+f_z}\\
0\\
\sqrt{1-f_z}\\
0
\end{pmatrix},\quad |f_z|\leq 1.
\end{align}
Also for this state $\langle\hat{F}_x\rangle=\langle\hat{F}_y\rangle=0$ and 
$\langle\hat{F}_z\rangle= nf_z$ and therefore 
the Hamiltonian and time evolution operator are given by Eqs. (\ref{eq:Hparallel}) and (\ref{eq:Uparallel}), respectively. 
The Bogoliubov matrix reads
\begin{align}
\label{eq:HB1m1}
\hat{B}_{1;-1} (t) =\hat{B}_{1;-1}^6(t)\oplus \hat{B}_{1;-1}^4. 
\end{align} 
Here $\hat{B}_{1;-1}^6(t)$ is time dependent with period $T=\pi/q$ and 
$\hat{B}_{1;-1}^4$ is time independent. The eigenvalues of $\hat{B}_{1;-1}^4$ are 
\begin{align}
\nonumber
\hbar\omega_{1,2,3,4} =&\pm\Big[\epsilon_k\Big(\epsilon_k+ (g_0+g_1)n\\ 
& \pm n\sqrt{(g_0-g_1)^2+4 g_0 g_1 f_z^2}\Big)\Big]^{1/2}.
\end{align}
Now $++,-+,+-$, and $--$ correspond to $\omega_1,\omega_2,\omega_3$, 
and $\omega_4$, respectively.  
These are all gapless modes. For rubidium the eigenvalues are real. 
In Appendix C we show that $\omega_{1,2}$ are density modes 
and $\omega_{3,4}$ are magnetization modes.  

We now turn to the eigenvalues of $\hat{B}_{1;-1}^6$. 
At $q=0$ $\hat{B}_{1;-1}^6$ becomes time independent and the eigenvalues are 
\begin{align}
\hbar\omega_{5,6} &= \pm\epsilon_k,\\
\nonumber
\hbar\omega_{7,8,9,10} &= \pm\frac{1}{\sqrt{2}}
\Big[2\epsilon_k^2+10\epsilon_k g_1 n +(g_1 n f_z)^2 \\
&\pm g_1 n \sqrt{(6\epsilon_k+g_1 n f_z^2)^2-8\epsilon_k f_z^2(4\epsilon_k-g_1 n)}\Big]^{1/2}. 
\end{align} 
For rubidium these are all real.
One of the eigenvalues $\hbar\omega_{4,5}$ has an energy gap $|g_1 n f_z|$. 
These eigenvalues describe spin-magnetization modes. 

\begin{figure}[h]
\begin{center}
\includegraphics[scale=.75]{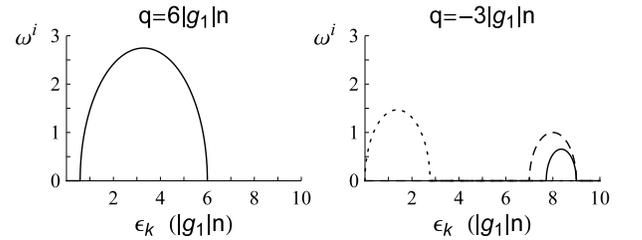} 
\end{center}
\caption{The positive imaginary part $\omega^\textrm{i}$ of the eigenvalues (\ref{omega1m1a}) 
and (\ref{omega1m1b}) related to $\psi_{1;-1}$ for $q=3$ and $q=-6$.   
The unit of $\omega^\textrm{i}$ is $|g_1|n/\hbar$. 
We have chosen $f_z=0$ as it gives the fastest-growing instabilities and the smallest size of a stable system.  
The solid and dashed lines correspond to $\omega_8^\textrm{i}$ and $\omega^{\textrm{i}}_5$, respectively, while the dotted line gives $\omega^{\textrm{i}}_7$ and $\omega^{\textrm{i}}_{10}$ 
[see Eqs. (\ref{omega1m1a}) and (\ref{omega1m1b})].
\label{fig:psi1m1fig}}
\end{figure}

For non-zero $q$ the stability can be analyzed using Floquet theory. 
As in the previous section, the fastest growing instabilities are obtained 
at $f_z=0$. This case can be studied analytically by changing basis as described in Appendix \ref{sec:appendixc}. The eigenvalues for the case $f_z=0$ are 
\begin{align} 
\label{omega1m1a}
&\hbar\omega_{5,6} = -3 q\pm \sqrt{(\epsilon_k+3 q)(\epsilon_k+ 2 g_1 n+3 q)},\\
\nonumber
\label{omega1m1b}
&\hbar\omega_{7,8,9,10} = 3q \pm \Big[(\epsilon_k+q)^2+4(\epsilon_k g_1 n+q^2)\\
&\pm 4\sqrt{[q^2+\epsilon_k (g_1 n+q)]^2-3 g_1 n q (\epsilon_k^2-q^2)}\Big]^{1/2}.
\end{align}
These are gapped excitations with a magnetic-field-dependent gap. In more detail, 
at $\epsilon_k=0$ we get $\hbar\omega_{5,6}=-3q\pm\sqrt{3q(2 g_1 n+3q)}$ and 
$\hbar\omega_{7,8,9,10}=3q\pm\sqrt{5q^2\pm 4q\sqrt{q(3 g_1 n+q)}}$. 
For positive $q$, the fastest-growing instability 
is determined by $\omega_8^{\textrm{i}}$ and is located approximately at $\epsilon_k =\textrm{max}\{0,-3+q\}.$ For negative $q$ there are three local maxima for $\omega^{\textrm{i}}$. The one with the largest amplitude is given by 
 $\omega^{\textrm{i}}_{7}$ and  $\omega^{\textrm{i}}_{10}$  and is located at $\epsilon_k\approx \textrm{max}\{0,q^2(|q|-1)/(q^2+|q|+1)\}$. The second largest is given by 
 $\omega^{\textrm{i}}_5$ and is at  $\epsilon_k\approx \textrm{max}\{0,3|q|+1\}$. Finally, the instability with the smallest amplitude is related to  $\omega^{\textrm{i}}_8$ and is at 
$\epsilon_k\approx \textrm{max}\{0,29 (10 |q|-1)/100\}$. In Fig. \ref{fig:psi1m1fig} 
we plot the behavior of $\omega^\textrm{i}$ for $q=6$ and $q=-3$. 
From Eqs. (\ref{omega1m1a}) and (\ref{omega1m1b}) it can be seen (see also Fig. \ref{fig:psi1m1fig}) 
that the state is stable if the size of the condensate is smaller than 
\begin{align}
\label{lambda1m1}
\lambda_{1;-1} =\frac{2\pi\hbar}{\sqrt{2m(2|q|-q)}}.
\end{align}

\subsection{Nonzero $\psi_2$ and $\psi_{0}$}
As the final example we consider a state  
\begin{align}
\psi_{2;0}=
\sqrt{\frac{n}{2}}\begin{pmatrix}
\sqrt{f_z}\\
0\\
\sqrt{2-f_z}\\
0\\
0
\end{pmatrix},\quad 0\leq f_z\leq 2.
\end{align}
As for other states considered in this article, now $\langle\hat{F}_x\rangle=\langle\hat{F}_y\rangle=0$ and 
$\langle\hat{F}_z\rangle= nf_z$ and 
the Hamiltonian and time evolution operator are given by Eqs. (\ref{eq:Hparallel}) and (\ref{eq:Uparallel}), respectively. We note that the stability properties of the states $\psi_{2;0}$ and 
$\psi_{0;-2}=\sqrt{n}(0,0,\sqrt{2-f_z},0,\sqrt{f_z})/\sqrt{2}$ are similar. Therefore the latter 
state will not be discussed in more detail. 
The Bogoliubov matrix of $\psi_{2;0}$ reads
\begin{align}
\label{eq:HB20}
\hat{B}_{2;0} (t) =\hat{B}_{2;0}^2\oplus\hat{B}_{2;0}^4\oplus \hat{B}_{2;0}^{4'}(t),
\end{align} 
where only $\hat{B}_{2;0}^{4'}$ is time dependent (with period $T=\pi/q$). The eigenvalues of $\hat{B}_{0;2}^2$ and $\hat{B}_{0;2}^4$ are 
\begin{align}
\hbar\omega_{1,2} = &\pm \epsilon_k, \\
\nonumber
\hbar\omega_{3,4,5,6} =& \pm\Big[\epsilon_k^2+ \epsilon_k( g_0 n+2 g_1 n f_z)\\
&\pm \epsilon_k n\sqrt{(g_0-2 g_1 f_z)^2+4 g_0 g_1 f_z^2} \Big]^{1/2}.
\end{align}
In the lower equation, $++, -+, +-$, and $--$ correspond to $\omega_3,\omega_4,\omega_5$, and $\omega_6$, respectively. 
These are gapless excitations. In Appendix \ref{sec:appendixd} we show that $\omega_{3,4}$ correspond to density modes, while $\omega_{1,2,5,6}$ are magnetization modes. For rubidium, these are all stable modes. 

\begin{figure}[h]
\begin{center}
\includegraphics[scale=.65]{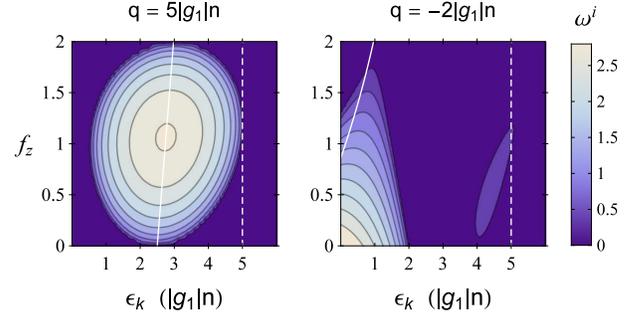} 
\end{center}
\caption{(Color online) The positive imaginary part $\omega^\textrm{i}$ of the eigenvalues 
(\ref{omega20}) related to $\psi_{2;0}$ for $q=5$ and $q=-2$.   
The unit of $\omega^\textrm{i}$ is $|g_1|n/\hbar$. The solid white line gives the approximate location of the fastest-growing instability, and the dashed white line corresponds to the largest possible size of a 
stable condensate, see Table (\ref{table}).
\label{fig:psi20fig}}
\end{figure}

After a suitable change of basis the Bogoliubov matrix $\hat{B}_{2;0}^{4'}$ becomes 
time independent, see Appendix \ref{sec:appendixd}. The eigenvalues of the new matrix are found to be  
\begin{align}
\label{omega20}
&\hbar\omega_{7,8,9,10} = \pm\frac{1}{\sqrt{2}}\sqrt{s_1 \pm \sqrt{(2\epsilon_k+g_1 n f_z +2 q)s_2}},
\end{align}
where
\begin{align}
\nonumber 
&s_1 = 2 \epsilon_k^2+(g_1 n f_z)^2 +4 \epsilon_k [(3-f_z)g_1 n +q]\\
\nonumber 
&\hspace{0.7cm} - 8 f_z g_1 n q +2 q (6 g_1 n+5q),\\
\nonumber 
& s_2 = f_z (g_1 n)^2[24(\epsilon_k+q)-10 \epsilon_k f_z -18 q f_z+ g_1 n f_z^2]\\
\nonumber
&\hspace{0.7cm} +32 q^2 [q+\epsilon_k +3 g_1 n (2- f_z)]  - 16 g_1 n q \epsilon_k f_z.
\end{align}
Now $++,-+,+-$, and $--$ are related to $\omega_7,\omega_8,\omega_9$, 
and $\omega_{10}$, respectively. These are gapped excitations and correspond to spin-magnetization modes. 
These modes can be unstable for rubidium; an example of the behavior of the positive imaginary component of  
$\omega_{7,8,9,10}$ is shown in Fig. \ref{fig:psi20fig}. 

An upper bound for the size of a stable condensate is the same as in the case of $\psi_{1;-1}$, see 
Eq. (\ref{lambda1m1}). With the help of Eq. (\ref{omega20}) it can be seen that the fastest-growing 
instability is approximately at $\epsilon_k=\textrm{max}\{0,-2+0.9 q+0.04 f_z (1+q)\}$ when $q>0$ and 
at  $\epsilon_k=\textrm{max}\{0,|q|-3+1.3 f_z-0.16 f_z^2\}$ when $q<0$.

\begin{table}[h]
\begin{tabular}{|c|c|c|c|}
\hline
State & $q$ & Stable size & Fastest-growing instability ($\epsilon_k$)\\
\hline
\hline
$\psi_{2;-1},$ & $>0$ & $\frac{2\pi\hbar}{\sqrt{4mq}}$ & $2q-g_1 n$\\
\cline{2-4}
$\psi_{1;-2}$ &$<0$  & $\frac{2\pi\hbar}{\sqrt{2m|q|}}$ & $|q|-\frac{5}{6}(2\mp f_z)|g_1|n$\\
\hline
$\psi_{2;0;-2}$ & $>0$ & $\frac{2\pi\hbar}{\sqrt{6mq}}$ &  $3q-g_1 n$\\
\cline{2-4}
  & $<0$ & $\infty$ & -\\
\hline
$\psi_{1;-1}$ & $>0$ & $\frac{2\pi\hbar}{\sqrt{2mq}}$ & $q-3g_1 n$\\
\cline{2-4}
 & $<0$ & $\frac{2\pi\hbar}{\sqrt{6m|q|}}$ & $\frac{q^2(|q|-g_1 n)}{q^2+|q| g_1 n+(g_1 n)^2}$\\
\hline
$\psi_{2;0},$ & $>0$ & $\frac{2\pi\hbar}{\sqrt{2mq}}$ & $-2+0.9 q+0.04 |f_z| (1+q)$\\
\cline{2-4}
$\psi_{0;-2}$ & $<0$ & $\frac{2\pi\hbar}{\sqrt{6m|q|}}$ & $|q|-3+1.3 |f_z| -0.16 f_z^2$\\
\hline
\end{tabular}
\caption{Summary of the results. Stable size gives the largest possible size of a stable 
homogeneous condensate and the fastest-growing instability indicates the approximate value of $\epsilon_k$ corresponding to the fastest growing instability. If $q$ is such that the $\epsilon_k$ given in the table 
is negative, the fastest-growing instability is at $\epsilon_k=0$.  On the second line of the table, 
the $-$ sign holds for $\psi_{2;-1}$ and the $+$ sign for $\psi_{1;-2}$. 
\label{table}}
\end{table}

\section{Conclusions} \label{sec:conclusions} 
In this article, we have studied the dynamical
stability of some nonstationary states of homogeneous $F=2$ rubidium  
BECs. The states were chosen to be such 
that the spin vector remains parallel to the magnetic field throughout the 
time evolution, making it possible to study the stability analytically. 
The stability analysis was done using the Bogoliubov approach in a frame of reference where the states were stationary. The states considered had two or three spin components populated simultaneously. These types of states were found to be stable in a rubidium condensate in the absence of a magnetic field, but a finite magnetic field makes them unstable. The wavelength and the growth rate of the instabilities depends on the strength of the magnetic field. The locations of the fastest-growing instabilities and the upper bounds for the sizes of stable condensates are given in Table \ref{table}. 
For positive $q$, the most unstable state, in the sense that its upper bound for the size of a stable condensate is the smallest, is $\psi_{2;0;-2}$. 
However, this is the only state that is stable when $q$ is negative. For $q<0$, the states giving the smallest size of a stable condensate are $\psi_{1;-1}$ and $\psi_{2;0}$. 

In comparison with $F=1$ condensates, the structure of the instabilities is much richer in an $F=2$ 
condensate. In an $F=1$ system, there is only one type of a state whose spin is parallel to the magnetic field. 
The excitations related to this state can be classified into spin and magnetization excitations \cite{Makela11}. 
In the present system, there are many types of states which are parallel to the magnetic field;  
we have discussed six of these. In addition to the spin and magnetization excitations, there 
exist also modes which change spin and magnetization simultaneously. The increase in the complexity can 
be attributed to the number of components of the spin vector. 

The stability properties of the states discussed in this article can be studied experimentally straightforwardly.   
These states had two or three non-zero components, a situation which can be readily achieved by current experimental means  \cite{Ramanathan11}. 
Furthermore, the stability of these states does not depend on the relative phases of the populated components, 
making it unnecessary to prepare states with specific relative phases.

Finally, we note that the lifetime of an $F=2$ 
rubidium condensate is limited by hyperfine changing collisions \cite{Schmaljohann04}. Consequently, the instabilities are visible 
only if the their growth rate is large enough compared to the lifetime of the condensate. We also remark that  
the stability analysis was performed for a homogeneous condensate, whereas in experiments an inhomogeneous trapping potential is used. The stability properties can be sensitive to the shape of this potential \cite{Klempt09}.

\appendix 
\section{Eigenvectors of $\hat{B}_{2;-1}$}\label{sec:appendixa}
Here we give the (unnormalized) eigenvectors of $\Hfour$, $\Hthree$, and $\Hthreep$. 
Unlike $\Hfour$, the operators $\Hthree$ and $\Hthreep$ depend on the magnetic field.  
The eigenvectors of the latter two are given at $q=0$. 
The operators $\Hfour$, $\Hthree$, and $\Hthreep$ will not be given here explicitly 
as they can obtained straightforwardly from Eqs. (\ref{X}) and (\ref{Y}). However, 
we give the bases with respect to which these operators and their eigenvectors are defined. 
The matrix $\Hfour$ is given in the basis 
$\{\mathbf{b}_1,\mathbf{b}_4,\mathbf{b}_6,\mathbf{b}_9\}$, where $\mathbf{b}_j$ 
is a ten-component vector with the $j$:th element equal to one 
and all other elements equal to zero. The eigenvectors of $\Hfour$ are 
\begin{align}
\mathbf{x}_j &=((\epsilon_k +\hbar\omega_j)\alpha_j,
(\epsilon_k +\hbar\omega_j),(\epsilon_k -\hbar\omega_j)\alpha_j,\epsilon_k -\hbar\omega_j),
\end{align}
where $j=1,2,3,4$ and 
\begin{align}
\alpha_j &\equiv \sqrt{\frac{1+f_z}{2-f_z}}\frac{g_0+4 g_1}{g_0-2 g_1}
\left(1+\frac{6 \epsilon_k g_0 g_1 n(2-f_z)}{(g_0+4 g_1)[\epsilon_k^2-(\hbar\omega_j)^2]}\right). 
\end{align} 
The matrix $\Hthree$ is defined in the basis $\{\mathbf{b}_2,\mathbf{b}_5,\mathbf{b}_8\}$.
At $q=0$ the eigenvectors are 
\begin{align}
\mathbf{x}_5 =&(\sqrt{2-f_z},-\sqrt{1+f_z},0),\\
\nonumber
\mathbf{x}_j =&(g_1 n\sqrt{(2-f_z)(1+f_z)},g_1 n (2-f_z),\\
&-\sqrt{\frac{3}{2}}(\epsilon_k+2 g_1 n-\hbar\omega_j)),\quad j=6,7.
\end{align}

By defining $\Hthreep$ with respect to the basis 
$\{\mathbf{b}_7,\mathbf{b}_{10},\mathbf{b}_3\}$ we get $\Hthreep=-(\Hthree)^*$. 
Therefore, the eigenvectors of $\Hthreep$ can be obtained from those of $\Hthree$ by complex conjugating. 

\section{Eigenvectors of $\hat{B}_{2;0;-2}$}\label{sec:appendixb}
The operator $\hat{B}_{2;0;-2}^6$ appearing in Eq. (\ref{eq:HB20m2}) 
is given in the basis $\{\mathbf{b}_1,\mathbf{b}_3,\mathbf{b}_5,\mathbf{b}_{6},\mathbf{b}_{8},\mathbf{b}_{10}\}$.   
The eigenvectors of $\hat{B}_{2;0;-2}^6$ corresponding to $\omega_{1,2}$ are
\begin{align}
\nonumber 
\mathbf{x}_1 =& (\sqrt{\rho_0 \rho_{-2}},-e^{i \theta}\sqrt{\rho_2\rho_{-2}},\sqrt{\rho_0\rho_2},0,0,0),\\
\mathbf{x}_2 =& (0,0,0,\sqrt{\rho_0 \rho_{-2}},-e^{i \theta}\sqrt{\rho_2\rho_{-2}},\sqrt{\rho_0\rho_2}). 
\end{align}
These are magnetization modes as they change the magnetization but not the spin direction. 
The exact eigenvectors corresponding to $\omega_{3,4,5,6}$ are too complicated to be given here. Therefore we approximate $g_1\approx 0$ (for rubidium $g_1/g_0\approx 0.01$)  and obtain   
\begin{align}
\nonumber 
\mathbf{x}_{3,4} \approx & (\sqrt{\rho_2},e^{i \theta}\sqrt{\rho_0},\sqrt{\rho_{-2}},\\
&-\gamma_\pm \sqrt{\rho_{2}},-\gamma_\pm \, e^{-i \theta} \sqrt{\rho_0},
-\gamma_\pm \sqrt{\rho_{-2}}),\\
\mathbf{x}_5 \approx & [(2-f_z)\sqrt{\rho_2},-e^{i\theta} f_z\sqrt{\rho_{0}},
-(2+f_z)\sqrt{\rho_{-2}},0,0,0],\\
\mathbf{x}_6 \approx & [0,0,0,(2-f_z)\sqrt{\rho_2},-e^{i\theta} f_z\sqrt{\rho_{0}},
-(2+f_z)\sqrt{\rho_{-2}}],
\end{align}
where
\begin{align}
\rho_{\pm 2} = & \frac{1}{4}(2-2 \rho_0 \pm f_z),\\
\label{gammapm}
\gamma_\pm = & \frac{1}{g_0 n}
\left[\epsilon_k+g_0 n\pm \sqrt{\epsilon_k (\epsilon_k+2 g_0 n)}\right]. 
\end{align}
Of these $\mathbf{x}_{3,4}$ are density modes and $\mathbf{x}_{5,6}$ are magnetization modes.

The operator $\hat{B}_{2;0;-2}^4$ is given in the basis $\{\mathbf{b}_2,\mathbf{b}_4,\mathbf{b}_7,\mathbf{b}_9\}$. 
$\hat{B}_{2;0;-2}^4$ is time dependent, but at $\rho_0=0$ the 
time evolution determined by $\hat{B}_{2;0;-2}^4$ can be solved analytically. 
With the help of the unitary basis transformation
\begin{align}
V = \frac{1}{\sqrt{2}}
\begin{pmatrix}
 e^{-3 i t q} & e^{-3 i t q}  & 0 & 0 \\
0 & 0 & e^{-3 i t q}  & e^{-3 i t q}  \\
0 & 0 & e^{3 i t q} & -e^{3 i t q}\\
e^{3 i t q} & -e^{3 i t q} & 0 & 0
\end{pmatrix},
\end{align}
we obtain a new Bogoliubov operator 
\begin{align}
\hat{\bar{B}}_{2;0;-2}^4\Big|_{\rho_0=0} 
\equiv V^\dag \hat{B}_{2,0;-2}^4\Big|_{\rho_0=0}  V +i\hbar\left(\frac{\partial}{\partial t} V^\dag\right)V,  
\end{align}
which is time independent. The eigenvectors of this operator are 
\begin{align}
\nonumber
\mathbf{x}_{7,8}  = &(2\hbar\omega_{4,5} -g_1 n f_z-6 q,\\
& 2\epsilon_k+[2+\sqrt{4-f_z^2}] g_1 n-6 q,0,0),\\
\nonumber
\mathbf{x}_{9,10}  = &(0,0,2\hbar\omega_{6,7} + g_1 n f_z -6 q,\\
& 2\epsilon_k+[2+\sqrt{4-f_z^2}] g_1 n-6 q). 
\end{align}
These modes change both magnetization and spin direction. 

\section{Eigenvectors of $\hat{B}_{1;-1}$}\label{sec:appendixc}
The operator $\hat{B}_{1;-1}^4$ is defined in the basis 
$\{\mathbf{b}_2,\mathbf{b}_4,\mathbf{b}_7,\mathbf{b}_9\}$. The  
eigenvectors are (in the limit $g_1=0$)
\begin{align}
%\nonumber
\mathbf{x}_1 &=(\sqrt{1+f_z},\sqrt{1-f_z},
-\gamma_-\sqrt{1+f_z},-\gamma_-\sqrt{1-f_z}),\\
\mathbf{x}_2 &=(-\gamma_- \sqrt{1+f_z},-\gamma_- \sqrt{1-f_z},
\sqrt{1+f_z},\sqrt{1-f_z}),\\
\mathbf{x}_3 &=(\sqrt{1-f_z},-\sqrt{1+f_z},0,0),\\
\mathbf{x}_4 &=(0,0,\sqrt{1-f_z},-\sqrt{1+f_z}).
\end{align} 
Here $\gamma_\pm$ is defined as in Eq. (\ref{gammapm}). 
Clearly $\mathbf{x}_{1,2}$ are density modes and $\mathbf{x}_{3,4}$ are magnetization modes. 

The operator  $\hat{B}_{2;0;-2}^6$ is defined in the basis 
$\{\mathbf{b}_1,\mathbf{b}_3,\mathbf{b}_5,\mathbf{b}_6,\mathbf{b}_8,\mathbf{b}_{10}\}$. 
Here we give the eigenvectors at $f_z=0$.
\begin{align}
\mathbf{x}_{5,6} &= (\epsilon_k+ g_1 n-\hbar\omega_{5,6}, g_1 n,0,0,0,0),\\
\mathbf{x}_{j} &= (0,0,\alpha_j,\beta_j,\gamma_j,\delta_j),\quad j=7,8,9,10,
\end{align}
where $\alpha_j,\beta_j,\gamma_j,\delta_j$ are too complex to be given here. 
These modes change both spin direction and magnetization.

\section{Eigenvectors of $\hat{B}_{2;0}$}\label{sec:appendixd}
The operators $\hat{B}_{2;0}^2,\hat{B}_{2;0}^4$, and $\hat{B}_{2;0}^{4'}$ are defined in the bases  
$\{\mathbf{b}_5,\mathbf{b}_{10}\}$,$\{\mathbf{b}_1,\mathbf{b}_3,\mathbf{b}_6,\mathbf{b}_8\}$, 
and $\{\mathbf{b}_2,\mathbf{b}_4,\mathbf{b}_7,\mathbf{b}_9\}$, respectively. 
The eigenvectors of $\hat{B}_{2;0}^2$ and $\hat{B}_{2;0}^4$  read 
\begin{align}
\mathbf{x}_{1,2} =& \mathbf{b}_{5,10},\\
\mathbf{x}_{3,4} = &(-\gamma_\pm\sqrt{f_z},-\gamma_\pm\sqrt{2-f_z},\sqrt{f_z},\sqrt{2-f_z}), \\
\mathbf{x}_{5} = & (\sqrt{2-f_z},\sqrt{f_z},0,0),\\
\mathbf{x}_{6} = & (0,0,\sqrt{2-f_z},\sqrt{f_z}).
\end{align}
Here  $\gamma_\pm$ is defined as in Eq. (\ref{gammapm}) and 
the index $3$ corresponds to $\gamma_+$ and the index $4$ to $\gamma_-$.  
Of these $\mathbf{x}_{3,4}$ are density modes  
and $\mathbf{x}_{1,2,5,6}$ are magnetization modes. 
When calculating $\mathbf{x}_{3,4,5,6}$ 
we have set $g_1=0$. 

The time dependence of the operator $\hat{B}_{2;0}^{4'}$ can be eliminated with 
the help of the basis transformation 
\begin{align}
V = \frac{1}{\sqrt{2}}
\begin{pmatrix}
 0 & 0 & e^{-i t q}  & 0  \\
0 & 0 & 0 & e^{- i t q}  \\
e^{i t q} & 0 & 0 & 0\\
0 & e^{-3 i t q} & 0 & 0
\end{pmatrix}
\end{align}
The eigenvectors $\mathbf{x}_{7,8,9,10}$ of the resulting time-independent operator 
describe spin-density modes and are too complicated to be given here.

\end{document}